\documentstyle[prl,multicol,aps,epsf]{revtex}

\begin{document}
\draft

\author{
Mikhail Titov$\P$, and Yan V. Fyodorov$\S,\P$,
}

\title{Time delay correlations and resonances in 1D disordered
systems}

\address{
$\P$  Petersburg Nuclear Physics Institute, Gatchina
188350, Russia }

\address{
$\S$ Fachbereich Physik, Universit\"at-GH Essen,
D-45117 Essen, Germany }

\date{\today}

\maketitle

\begin{abstract}
The frequency dependent time delay correlation
function $K(\Omega)$ is studied analytically
for a particle reflected from a {\it finite} one-dimensional
disordered system.
In the long sample limit $K(\Omega)$ can be used to extract
the resonance width distribution $\rho(\Gamma)$.
Both quantities are found to decay algebraically  as
$\Gamma^{-\nu}$, and $\Omega^{-\nu}$, $\nu\simeq 1.25$
in a large range of arguments.
 Numerical
calculations for the resonance width distribution in 1D
non-Hermitian tight-binding
 model agree reasonably with the analytical formulas.
\end{abstract}

\begin{multicols}{2} \narrowtext

For a long time one dimensional disordered conductors set an
example for understanding the localization phenomenon
in real electronic systems. In this communication we consider the
reflection of a spinless particle from the disordered region
of length $L$ and address such quantities as the resonance
widths distribution and the time delay correlation function.

The issue of time delays and resonances attracted considerable
attention in the domain of chaotic scattering\cite{YS}
 and mesoscopics
because of relevance for properties of small metallic granulas
and quantum dots \cite{BM}. An essential
 progress in understanding of statistical fluctuations of these
and related quantities
turned out to be possible recently \cite{YS,BM,Sok,YS1}
in the context of random matrix description
of chaotic motion of quantum particles in the scattering region.
Such a description assumes from the very beginning that
the time of relaxation on the energy shell due to
scattering on impurities (Thouless time) is negligible in
comparison with the typical inverse spacing between neighboring
energy levels (Heisenberg time). This assumption neglects
effects of Anderson localization completely.
At the same time the localized states are expected to effect
the form of the resonance width statistics and time delay
correlations considerably.

Difficulties in dealing with the
localization analytically in higher dimensions
suggest one-dimensional samples with the white noise random
potential as a natural object of study.
Various aspects of particle scattering in such systems were
addressed by different groups recently \cite{kumar,CT,MK}.
Still, the statistical properties of resonances in such systems
were not studied to our best knowledge.

It is well-known that the level spacing statistics becomes Poissonian
with onset of the Anderson localization
when the system size $L$ far exceeds the localization length
$\xi$ and eigenstates do not overlap any longer.
Notwithstanding the fact that
the argumentation is applied to a closed system
the same should be true for its open counterpart as well.
To exploit this fact we define the two-dimensional
spectral density ${\tilde \rho}(Z)$ associated with
broadened levels (resonances)
${\tilde \rho}(Z)=\sum_n\delta(X-ReZ_n)\delta(Y+ImZ_n)$,
where $Z=X+iY$ is the complex energy, $Z_n=E_n-i\Gamma_n/2$
is the coresponding position of the $n$-th resonance whose width is
$\Gamma_n$. We further define the resonance
correlation function as follows:
%
%
        \begin{equation}
        \label{1}
        \langle {\tilde \rho}(Z_1){\tilde \rho}(Z_2)
        \rangle_c = \langle {\tilde \rho}(Z_1)
        \rangle \delta^{(2)}(Z_1-Z_2)- {\cal Y}_2(Z_1, Z_2)
        \end{equation}
with the brackets standing for the averaging over disorder.
The so-called cluster function ${\cal Y}_2(Z_1, Z_2)$ reflects
eigenvalue correlations and therefore
should vanish in the thermodynamic limit $L\to \infty$.
This property provides us with a possibility to relate the averaged density of
 resonances in the complex plane $\langle \rho(Z)\rangle$ to the time delay
correlation function.

We recall that the time delay at a fixed real energy $E$ can be written
in terms of ${\tilde \rho}(X, Y)$ as (see e.g. \cite{YS,YS1})
%
%
        \begin{equation}
        \label{2}
        \tau(E)=\int_{0}^{\infty}\!\!\!\!dX
        \int_{0}^{\infty}\!\!\!\!dY
        \frac{2Y {\tilde \rho}(X, Y)}{(E-X)^2+Y^2}
        \end{equation}
As far as we restrict ourselves to the thermodynamic limit
the correlation of time delays at two different energies
$E+\Omega$, and $E-\Omega$ can be expressed in terms of the
correlation function Eq.(\ref{1}) with the cluster function
neglected. The formula obtained can be further simplified
employing the integration over $X$ and taking into account
that semiclassically $\Omega, Y \ll E$.
In addition we find it to be convenient to rescale the
spectral density and introduce the dimensionless resonance
widths distribution function $\rho(y)$ by means of the relation
%
%
        \begin{equation}
        \label{dos3}
        \langle{\tilde \rho}(X,Y)\rangle
        = 2\pi\nu^2_{\xi}(X) \rho(2\pi \nu_{\xi}(X)\ Y)
        \end{equation}
where $\nu_{\xi}(E)\simeq \xi (2 \pi \sqrt{E})^{-1}$ is the density
of states corresponding to the system of the size $\xi$.
This is done, the relation between the time delay correlation function and
the widths distribution takes the form:
%
%
        \begin{equation}
        \label{4}
        \begin{array}{rcl}
        K(\omega, \frac{L}{\xi})&=& \Delta_{\xi}^2
        \langle \tau (E+\omega\Delta_{\xi})
        \tau (E-\omega\Delta_{\xi})
        \rangle_c\\ &=&
        \int_{0}^{\infty}dy
        \frac{y}{\omega^2+y^2} \rho(y)
        \end{array}
        \end{equation}
where $\Delta_{\xi}=(2 \pi \nu_{\xi})^{-1}$ standing for the mean level
spacing corresponding to a single localization volume is a
natural energy scale for measuring the resonance width as well as the
time delay correlations. As we will see below
Eq.(\ref{4}) gives us a possibility to
restore the density $\rho(y)$ from the time delay correlation function.

Before we proceed with our analysis
it is instructive to develop a simple intuitive picture
about the form of the resonance width distribution
for a particle reflection from one dimensional disordered sample.
We may say qualitatively that this distribution (normalized
against the number of states N) is given by
%
%
        \begin{eqnarray}
        \label{naive1}
        {\cal P}(y)&=&\frac{N
        \xi}{L}p(y)+\frac{N}{L}\int_{\xi}^{L} dx\ \delta\left(
        y-e^{-x/\xi}\right) \\
        \label{naive2}
        &=&\frac{N\xi}{L}\left[p(y)+\frac{1}{y}\
        \chi\left( e^{-L/\xi}\le y \le e^{-1}\right)\right]
        \end{eqnarray}
where $y=\Gamma/\langle \Gamma\rangle$,
and , and $\chi(y)$ is a characteristic function of the interval.
The variable $x$ in Eq.(\ref{naive1}) stands for the distance from the
open edge, and we assume the opposite edge of the sample to be
closed for the sake of simplicity.
The function $p(y)$ is the resonance width distribution
corresponding to resonances residing at a distance $x\lesssim \xi$ from the
open edge.  Other states which are localized in the bulk
give rise to
exponentially small probability for the particle to escape
from the sample and the corresponding width is
$\Gamma\sim \exp{(-x/\xi)}$.
This is oversimplified picture can give us an
idea of $y^{-1}$ dependence of the resonance density in a rather wide
parametric range. However our explicit calculations presented below
demonstrate a slightly different power law: $y^{-1.25}$.


In the present communication we give only a
sketch of the calculations, relegating details to a more
extended publication. To start with, we consider an elastic
scattering of a spinless particle $\psi(x)$ on 1D disordered sample.
The boundary condition $\psi(0)=0$ is imposed at the origin
(left edge) and the particle is subject to the white-noise potential
$V(x)$, $\langle V(x) \rangle=0$, inside the interval $(0, L)$.  No
potential barrier is assumed at the right edge which corresponds to the
perfect coupling between the system and the continuum.  The
Shr\"odinger equation on the interval $(0,L)$ can be
formulated in terms of the logarithmic derivative of the wave function
$z=d\ln\psi/d x$ \cite{CT,LGP}:
%
%
        \begin{equation}
        \label{Schr1}
        \frac{dz}{dx}=V(x)-E-z^2
        \end{equation}
where $\langle V(x)V(y) \rangle=\sigma_g \delta(x-y)$, $E=k^2$.
In what follows we scale the variance
in such a way that  $\sigma_g=4E/\xi$, with $\xi$ being
the localization length and use the rescaled potential
$v(x)=V(x)/2k$ with the variance $1/\xi$.

After
the substitution $z=k\cot{\phi}$ Eq.(\ref{Schr1}) is reduced to the
equation on the phase $\phi$. The phase shift between in- and outgoing
waves can be checked to be equal to $2\phi$. Its derivative over the
energy is nothing but the corresponding time delay.
%
%
        \begin{equation}
        \label{tau1}
        \tau=\frac{1}{k} \frac{d\phi}{dk}
        \end{equation}
To address time delay correlations at two different
energies we have to take into account the set of the following four
equations (cf. \cite{CT,LGP}):
%
%
        \begin{equation}
        \label{Lang3}
        \begin{array}{rcl}
        {\dot \phi_1}&=&k+q-2v(x)\sin^2{\phi_1} \\
        {\dot \phi_2}&=&k-q-2v(x)\sin^2{\phi_2} \\
        {\dot \tau_1}&=&k^{-1}-2\tau_1 v(x) \sin{2\phi_1}\\
        {\dot \tau_2}&=&k^{-1}-2\tau_2 v(x) \sin{2\phi_2}
        \end{array}
        \end{equation}
where the dot states for the derivative over $x$.
The first two equations are those for phases each related
to its own eigenstate $E=(k+q)^2$, and $E=(k-q)^2$. Two more equations
are those for the time delays. They emerge
after taking the derivative of the corresponding
phases over the energy $E=k^2$ and exploiting
an approximation $q \ll k$. Eqs. (\ref{Lang3}) contain a multiplicative
random force $v(x)$
and must be understood in the Stratonovich sense \cite{G}.

As usual, significant simplifications occur in semiclassical limit
when we restrict ourselves to the large energies: $k \gg q$,
$kL \gg 1$ \cite{B}.
Last condition means that the joint probability density
$P(\phi_1, \phi_2, \tau_1, \tau_2)$ satisfying
a Fokker-Planck equation oscillates rapidly with respect to the
phase $\phi=\phi_1+\phi_2$. We can average the probability density
over this rapid phase \cite{LGP} in the leading approximation over
$1/kL$.
This averaging can be most efficiently done
on the level of the Fokker-Planck equation and, as the result,
we obtain a partial differential equation on the density
$P(\eta, \tau_1, \tau_2)$ where $\eta=\phi_1-\phi_2$ is a "slow" variable.
This is done, it is more convenient to come
back to the level of the Langevin-type equations for only three
variables $\eta$, $\tau_1$, and $\tau_2$.
Such a procedure requires
introducing of two independent random forces $f_1(x)$, $f_2(x)$
rather then one.
The reduced set of the Langevin type equations is found to be
%
%
        \begin{equation}
        \label{Lang4}
        \begin{array}{rcl}
        {\dot \eta}&=&2q-\xi^{-1}\sin{\eta}\cos{\eta}
        +\sin{\eta}f_1(x) \\
        {\dot \tau_1}&=&\frac{1}{k}+\tau_1(
        \cos{\eta}f_1(x)-\sin{\eta}f_2(x)
        -\frac{1}{\xi}\cos^2{\eta})\\
        {\dot \tau_2}&=&\frac{1}{k}+\tau_2(
        \cos{\eta}f_1(x)+\sin{\eta}f_2(x)
        -\frac{1}{\xi}\cos^2{\eta})
        \end{array}
        \end{equation}
The first equation is of the most importance and governs the
evolution of the slow phase $\eta$. It contains the only random
force $f_1$: $\langle f_1(x) f_1(y) \rangle = (2/\xi)\delta(x-y)$.
The other two equations depend on additional random force
$f_2$, which is
independent of $f_1$ and has the same variance $2/\xi$.
In the initial point $x=0$ the variables $\eta$, $\tau_1$, and
$\tau_2$ have to be equal zero. The value of
the parameter $q$ influences the dynamics of the phase $\eta$:
when $q$ is positive (negative) the function
$\eta(x)$ is also positive (negative).
The limit $q=0$ leads to the trivial stationary
solution $\eta(x)=0$ and two
equations for $\tau_1$, $\tau_2$ become equivalent and
can be used for calculating the time delay distribution
\cite{CT}.
In a general case $q \neq 0$ the last two equations still
can be solved:
%
%
        \begin{equation}
        \label{tau2}
        \tau_{1,2}(L)=\frac{1}{k}\int\limits_{0}^{L}\!
        \!dx\>
        e^{\int_{x}^{L}dx'(f_1\cos{\eta}
        \mp f_2\sin{\eta} -\frac{1}{\xi}\cos^2{\eta})}
        \end{equation}
To perform the disorder averaging we have to integrate
over the random forces $f_1$, $f_2$
with the Gaussian measure. When $q=0$ the averaging
of the product $\tau_1 \tau_2$
is readily done and leads to the
second moment of the time delay distribution:
%
%
        \begin{equation}
        \label{tau3}
        \langle \tau^2 \rangle =\frac{1}{2 \Delta_{\xi}^2}
        \left(e^{2T}-2T-1\right); \qquad T=\frac{L}{\xi}
        \end{equation}
which increases exponentially with the length of the sample.
It is even simpler to check that $\langle \tau \rangle= L/k
\simeq \Delta_{\xi}T$.
When $q \neq 0$ the averaging of the product $\tau_1\tau_2$
over the random forces gives rise to
the correlation function (\ref{4})
where $\omega=2q \xi$. The integration over the
noise $f_2$ is still Gaussian and can be easily done.
To perform the other integration we restrict our consideration
to the interval $\eta \in (0,\pi)$ because of the obvious
periodicity of the solutions and take advantage of the
new variable $g=\ln{(\tan{\eta/2})}$ to rewrite
the first of Eqs.(\ref{Lang4}) in the following form:
%
%
        \begin{equation}
        \label{trans}
        f_1={\dot g}-V(g); \qquad
        V(g)=2q\cosh{g}+\xi^{-1}\tanh{g}
        \end{equation}
This gives a possibility to replace
the integration over $f_1$ by that over $g$ taking into
account the initial condition $g(0)=-\infty$  and the corresponding
Jacobian:
%
%
        \begin{equation}
        \label{Jac}
        {\cal J}=
        \exp{\left\{\frac{1}{2}\int_{0}^{L}\frac{dV(g)}{dg}dx
        \right\}}
        \end{equation}
The next step is to make use of
the formal analogy with the Feynmann path integral of the
quantum mechanics.
After straightforward calculations we come back
to the variable $\eta$ and arrive at
the time delay correlation function (\ref{4}) in the
following form:
%
%
        \begin{equation}
        \label{td1}
        K(\omega, T)=2\int_{0}^{T}dt_1
        \int_{0}^{t_1}dt_2 R(T, t_1, t_2; \omega)-T^2
        \end{equation}
        \begin{equation}
        \label{td2}
        \begin{array}{l}
        R= \int_{0}^{\pi}d\eta\ \delta(\eta)
        e^{t_2 {\hat H}_L}
        e^{(t_1-t_2){\hat H}_C}
        e^{(T-t_1){\hat H}_R}\ 1\\
        {\hat H}_L=\sin^2{\eta}
        \frac{d^2}{d\eta^2}+\omega\frac{d}{d\eta},\\
        {\hat H}_C=\frac{d}{d\eta}\sin^2{\eta}\frac{d}{d\eta}
        +\omega\frac{d}{d\eta},\\
        {\hat H}_R=\frac{d^2}{d\eta^2}\sin^2{\eta}
        +\omega\frac{d}{d\eta}
        \end{array}
        \end{equation}
The action of the differential operators
$\exp{{\hat H}_{L,C,R}}$
 applied to the right-hand side is equivalent to finding
the solution of the corresponding evolution equations.
Unfortunately the eigenstates and eigenfunctions
of the operators
${\hat H}_{L,C,R}$ are poorly understood
and an exact calculation of the quantity $R$
for arbitrary sample length $L$ and frequency $\omega$
is beyond our possibilities.
Still, the approximate solution can be found.
It is worth noticing that the equations of this kind
are common in the context of one dimensional
localization and have been disscussed in the literature
\cite{B,M,GDP,AP,AS}.
The thorough analysis of Eqs.(\ref{td1},\ref{td2}) shows
that for exponentially small frequencies $\omega \ll \exp{(-T)}$
the main contribution to the integral over
$t_1$, $t_2$ in Eq.(\ref{td1}) comes from the region
$t_1 \sim t_2 \ll T$.
In this situation the time delay correlation function
is proportional to
$\lim_{\eta\to 0}\exp{(T {\hat H}_R)}1$. The latter
expression is appropriate for the perturbation analysis in $\omega$
which allows to present the result in the form of the asymptotic
series:
%
%
        \begin{equation}
        \label{small_o}
        K(\omega,T)=\sum_{k=0}^{\infty}\gamma_k(i\omega)^{2k}
        e^{(2k+1)(2k+2)T}
        \end{equation}
where $\gamma_k$ does not depend on $T$. With the help of the relation
(\ref{4}) we obtain \cite{CT,MK,AP}
that the resonance widths are distributed
according to the log-normal law in the region of
exponentially narrow resonances $y\ll \exp(-T)$:
%
%
        \begin{equation}
        \label{small_y}
        \rho(y)=N^{-1}y^{-3/2}\exp{\left(
        -(4T)^{-1}\ln^2{y}
        \right)}
        \end{equation}
in agreement with natural expectations \cite{CT,MK,AP}.

Eqs.(\ref{td1},\ref{td2}) can be analyzed also in the
thermodynamic limit $T\to \infty$ or large enough
frequencies $\exp(-T)\ll\omega$. The main contribution
to the integral Eq.(\ref{td1}) exactly cancels the term $T^2$
and the leading correction turns out to be independent on $T$.
%
%
        \begin{equation}
        \label{K}
        K(\omega)=-\frac{\pi^2}{2^8}\int\limits_{-\infty}^{\infty}
        d\mu\
        \frac{\mu^2 \Gamma\left(\frac{3+i \mu}{2}\right)
        \left(|\omega|/8\right)^{\frac{i\mu-3}{2}}}
        {\Gamma^2\left(1+\frac{i \mu}{2}\right)
        \cosh^2\frac{\pi\mu}{2}}
        \end{equation}
The result of the numerical evaluation of the present integral is
shown in Fig.1.

In the derivation of the expression above
we consider the operators ${\hat H}_{R,C,L}$ acting in the
Fourier space $\exp(2i m \eta)$.
The drastic simplifications
occur when we let the index $m$ to be continous which is
the correct approximation for $|m|\gg 1$. Such a limit should be
appropriately matched with that for small $m$:
$|m \omega|\ll 1$ \cite{AP,M,GDP}.
The continous approximation is responsible for an unphysical tail
for large frequencies. We expect, however,
that Eq.(\ref{K}) is valid for $\omega \lesssim 1$.
It is interesting to note that even though a simple
one-pole approximation of the integral (\ref{K}) gives rise to
$\omega^{-1}$ dependence in $\omega \to 0$ limit, such a dependence
takes place only for extremely small frequencies: $\omega \lesssim 10^{-8}$.
The best fit for the correlation function in a very broad
frequency region is $\omega^{-1.25}$.
Let us derive now the resonance width distribution $\rho(y)$
which follows from Eqs.(\ref{4},\ref{K}). Taking into account
the identity
%
%
        \begin{equation}
        \label{restore}
        \int_{0}^{\infty}\!\!\!\!dy\ y^{\alpha+1}
        (y^2+\omega^2)^{-1}=
        -\pi |\omega|^{\alpha}/(2\sin{\pi\alpha/2})
        \end{equation}
we restore the resonance density in the following form:
%
%
        \begin{equation}
        \label{rho}
        \rho(y)=\frac{\pi}{2^7}\int\limits_{-\infty}^{\infty}
        d\mu\
        \frac{\mu^2\Gamma\left(\frac{3+i \mu}{2}\right)
        \sin{\frac{\pi(i\mu-3)}{4}}(y/8)^{\frac{i\mu-3}{2}}}
        {\Gamma^2\left(1+\frac{i\mu}{2}\right)
        \cosh^2\frac{\pi\mu}{2}}
        \end{equation}
This integral shows basically the same features
as Eq.(\ref{K}), (see Fig.\ref{fig:res1}). The resonance widths
turn out to be virtually cut at $\Gamma\simeq \Delta_{\xi}/8$
and the rest part of the plot should not to be taken seriously
being an artifact of the approximation used.
\begin{figure}
\centerline{\epsfxsize=8cm
\epsfbox{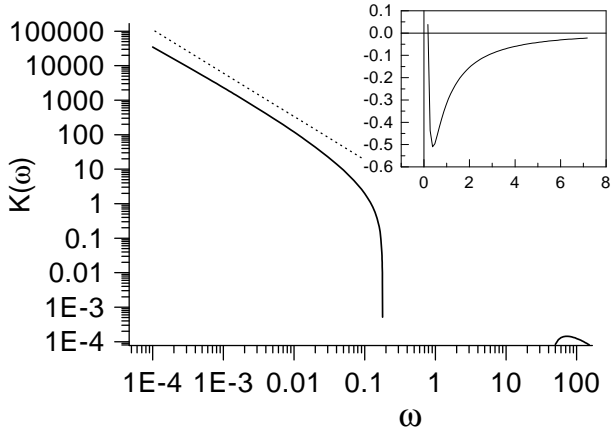}}
\caption{ The time delay correlation function as
given by Eq.(\ref{K}) (solid line). The negative part of the
function shown in the insert is the artifact of the
approximation done. The dotted line corresponds to
the dependence $\omega^{-1.25}$. The function dropes down
to zero at the point $\omega\simeq 1/8$.
 \label{fig:tdelay}  } \end{figure}
\begin{figure}
\centerline{\epsfxsize=8cm
\epsfbox{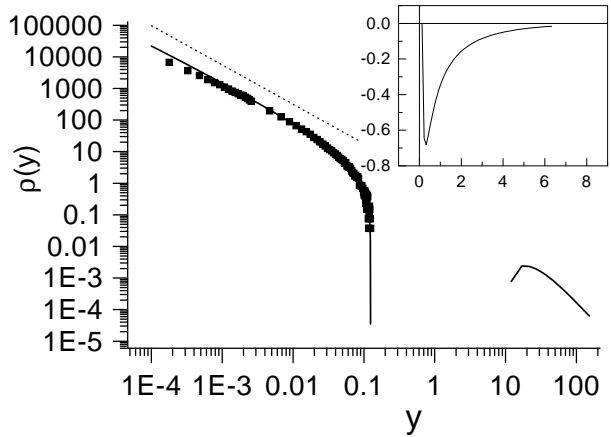}}
\caption{The resonance widths distribution calculated by
the use of Eq.(\ref{rho}) (solid line)
as compared with the results of the
numerical simulations (squares) done
for the matrices of the size $300*300$.
The insert and the dotted line are like
those in Fig.\ref{fig:tdelay}.
\label{fig:res1}  } \end{figure}
To check our results we considered
the simplest random matrix model which was expected to
belong to the same universality class. Namely we
diagonalize numerically as many as $10000$ tridiagonal
matricies of the size $300*300$ with its diagonal elements being
uniformly distributed in the interval $(-1,1)$.
The off-diagonal elements are chosen to be equal to $1$.
In the same manner as in \cite{YS} we can argue
 that an effect of the open edge can be effectively simulated
by adding the imaginary shift $\gamma=i\pi$ to last diagonal element
of the matrix.
We picked up eigenvalues from the center of the spectrum
and investigated statistics of their imaginary parts. As shown in
the Fig.\ref{fig:res1} the numerical results agree reasonably with
our analytical predictions.

In conclusion we address analytically the time delay correlations
and resonances for the problem of reflection from 1D disordered sample.
We find the resonance density $\rho(\Gamma)$ to reveal the
log-normal behaviour for the exponentially small widths
and the algebraic dependence
close to $\Gamma^{-1.25}$ in the wide parametric range
$\exp{(-L/\xi)}\ll \Gamma/\Delta_{\xi} \ll 1$. The time
delay correlations are found to demonstrate similar
behavior as a function of frequency.

We greatfully acknowledge very informative and stimulating discussions with
A.~Comtet and C.~Texier. MT appreciate A.~Comtet for kind hospitality
extended to him in Orsay, Paris, where the part of this work was done.
The work was supported by INTAS Grant No.~97-1342, (MT, YF), SFB 237
"Disorder and Large Fluctuations" (YF),
Russian Fund for Statistical Physics, Grant VIII-2,
and RFBR grant No 96-15-96775 (MT).


\end{multicols}

\end{document}